\documentclass[12pt]{article}

\usepackage{graphicx}
\usepackage{amssymb,amsmath,latexsym}
\usepackage[T1,hyphens]{url}
\usepackage[colorlinks,urlcolor=blue]{hyperref}

\begin{document}

\title{Spurious,
Emergent
Laws in Number Worlds}

\author{Cristian S. Calude \\
        {\small   School of Computer Science, University of Auckland,} \\
        {\small   Private Bag 92019, Auckland, New Zealand}  \\
%        {\small   email: cristian@cs.auckland.ac.nz, URL: http://www.cs.auckland.ac.nz/{\textasciitilde}cristian}\\
   {\small  \url{http://www.cs.auckland.ac.nz/~cristian}   }  \\
{\small }\\
        Karl Svozil \\
        {\small   Institute for Theoretical Physics, Vienna University of Technology,} \\
        {\small   Wiedner Hauptstrasse 8-10/136, 1040 Vienna, Austria}\\
%        {\small   email: svozil@tuwien.ac.at, URL: http://tph.tuwien.ac.at/{\textasciitilde}svozil}
{\small   \url{http://tph.tuwien.ac.at/~svozil}}
       }

\date{\today}

\maketitle

\thispagestyle{empty}
\begin{abstract}
We study some aspects of the emergence  of {\it l\'ogos} from {\it x\'aos} on a basal model of the universe using methods and techniques from algorithmic information and Ramsey theories. Thereby an intrinsic and unusual mixture of meaningful and  spurious, emerging laws surfaces. The~spurious, emergent  laws abound,  they can be found almost everywhere. In accord with the ancient Greek theogony one could say that {\it l\'ogos}, the Gods and the laws of the universe, originate from ``the void,'' or from {\it x\'aos}, a picture which supports  the unresolvable/irreducible lawless hypothesis. The~analysis presented in this paper suggests that the ``laws'' discovered in science correspond merely to syntactical correlations, are local and not universal.
\end{abstract}

\section{Introduction}
\label{2018-was-sec1}
What if the universe, on the most fundamental layer, just consisted of numbers?
This is a suspicion at least as old as the Pythagoreans.
According to Huffman's entry in {\it The Stanford Encyclopaedia of Philosophy}~\cite{sep-pythagoreanism},
``\dots  in the {\em Metaphysics}, he [Aristotle] treats most Pythagoreans
as adopting a mainstream system in contrast to another group of Pythagoreans
whose system is based on the table of opposites \dots .
The central thesis of the mainstream system is stated in two basic ways:
the Pythagoreans say that things are numbers or that they are made out of numbers.
In his most extended account of the system in {\em Metaphysics}~1.5,
Aristotle says that the Pythagoreans were led to this view
by noticing more similarities between things and numbers than between things and the elements,
such as fire and water, adopted by earlier thinkers.''
Moreover, according to another contemporary review~(\cite{mg1968},~p.~14),
``according to Aristotle, the Pythagoreans do not place the objects of mathematics
between the ideas and material things as Plato does,
they say `that things themselves are numbers' and that
`number is the matter of things as well as the form of their modifications and permanent states'.
As the principles of mathematics, numbers are the `principles of all existing things'.''

%`For them [Pythagoreans], the principle of all things is the monad; arising from the monad, the undetermined dyad acts as matter to the monad which is cause; from the monad and the undetermined dyad arise numbers; from numbers points; from these, lines out of which arise plane figures which produce in turn solid figures; from these, material bodies whose constituents are four --  fire, water, earth, air. These elements interchange and turn into another completely; out of them arises a world which is animate, intelligent, spherical and has the earth as its center, a spherical body inhabited round about."
%
The importance of number, and more generally, of mathematics,  for not only describing
but ``being'' the bricks of the universe was stressed  by eminent physicists, like
Schr\"odinger~(\cite[]{schroed:natgr}, Chapter~III).
%  ``The basic doctrine of the Pythagoreans, we are told,
%was that {\em things are numbers.}''
The~introduction of computing machinery creating virtual realities
brought these issues to the forefront~\cite{zuse-69,zuse-70,fredkin,toffoli:79,margolus-billard,wolfram-2002}.
In a recent bold leap,
Tegmark's {\em Mathematical Universe
Hypothesis}~\cite{tegmark2007,tegmark2014} states that ``the physical universe is not merely described by
mathematics, but is a mathematical structure''. As a consequence, mathematical existence equals physical existence, and
all structures that exist mathematically (even in a non-constructive way) exist physically as well.

How could things be numbers? A  world ``spanned'' by numbers can be represented by a single infinite (binary)  sequence\footnote{\it A sequence is infinite while a string is finite. A finite prefix of a sequence is then a string.},
or, equivalently,  a single real number.

In what follows a {\em number world}
will be  modelled by a (binary) sequence.\footnote{This physical-mathematical mapping assumption
is essential for this paper.}
Our choice is  not to operate with the  more geometric-centred Ancient Greek concept of number,
which is essential for  many continuous models of mathematical physics,
but with an  algorithmic one which is capable of giving a global perspective of the universe.
Adopting this framework is motivated by
Plato's mathematical discussion, in {\em Timaeus},
of the relations between numbers and things, see~\cite{mg1968},~p.~14 and also~\cite{calude2013theeinai},
and it is adopted here as a matter of {\em hypothesis}.

All entities encoded therein, including observers as well as measured objects, must be embedded
in~\cite{toffoli:79,svozil-94};
that is, they must themselves be (formed out of) numbers  or symbols~\cite{borges-library}.
Non-numeric properties associated with such a ``world on a sequence''
can arise by way of a structural, levelled~hierarchy~\cite{anderson:73}.

Epistemologically this can be perceived as
``emergence\footnote{``Emergence is a notorious philosophical term of art.'' ~\cite{sep-properties-emergent}.  In this paper we will not use the term in the sense of the philosophical emergency theory, but with the signification given in physics~\cite{Kivelson-16}:  {\it ``The term emergent is used to evoke collective behaviour of a
large number of microscopic constituents that is qualitatively
different than the behaviours of the individual constituents." } } of reality'', which is the inverse
of reductionism to some more fundamental, basic levels, involving explanations in terms of ever ``smaller'' entities:
physical/universal/natural laws---in particular, relational and probabilistic ones---arise  as effective patterns
and structures ``bottom-up'' (rather than ``top-down'').

Such concepts were quite popular in the  {\it fin de si\'ecle} Viennese physical circles, so much so
that they have been referred to as
the {\em Austrian Revolt in Classical Mechanics}~\cite{Hiebert2000} and
{\em Vienna Indeterminism}~\cite{Stoeltzner1999}: stimulated by
the apparent indeterminacy manifesting  in Rutherford's asymptotic decay law
and its corroboration  by Schweidler~\cite{schweidler-1905},
Exner's 1908 inaugural lecture as {\it Rector Magnificus}
included the suggestion that~(\cite[]{Exner-1908}, p.~18)  ``we have
to perceive all so-called exact laws as probabilistic which are not valid with
absolute certainty; but the more individual processes are involved the higher their certainty''.
Also Schr\"odinger's inaugural lecture in Z\"urich entitled {\em ``What is a natural law?''}
adopted and promoted Exner's ideas~\cite{schrodinger-1929,book:16081},
well in accord with Born's later inclinations~\cite{born-26-1}.
Since then classical statistical physics, as well as radioactive decay processes
and quantised systems have operated under the presumption that
the most fundamental layers of microphysical description
are---both theoretically as well as phenomenologically and empirically---consistent with irreducible indeterminism.

Later related ideas have been brought forward in the context of
a layered structure of physical theories~\cite{anderson:73},
emergent  cognition---perceived as an ``emerging epiphenomenon'' of neural activity;
not unlike traffic jams they arise  from the movements of individual taxis~\cite{hofstadter:82}
--as well as emergent computation~\cite{Stephani-1990}.

In what follows we shall,
in a ``Humean spirit''~\cite{Hume-Treatise},
study  ``laws'' as  patterns/correlations  in sequences using  the concept of {\em spurious
correlations in data}, to be defined  later.
Two guiding theories will be applied: one is algorithmic information theory, the other is Ramsey theory.
The gist of these two ways of looking at data is twofold: ``all very long, even irregular'' data sequences
contain ``very large'' (indeed, as long as you prefer) regular,
computable and thus, in physical terms, {\em deterministic}, subsequences.
Secondly,
it is impossible and inevitable for any arbitrary data set {\em not} to contain a variety of spurious correlations; that is,
relational properties which could physically be  wrongly  interpreted as
laws ``governing'' that universe of data.

\section{Physical/Universal/Natural Laws}%Natural law}

The notion of law in  natural sciences,
or law of  the universe~\cite{Beebee-Hume,sep-kant-hume-causality,Russell-1913-cause,mumford-causation,feynman-law,Norton-2003-cafs} has a long
ambivalent history.
It might not be overstated to claim that
the conjecture
that there are laws of nature is the core to what science is and how it was and is performed.
Of course, one can refute this view and this lawless hypothesis   has been discussed by various authors,
see~\cite{armstrong_1983,vanFraassen1989-VANLAS,calude1999islawful,lawlses_rosen2010,calude2013theeinai,chaos_multiverse2017,Mueller-2017,Cabello-2018-BornRule}.
Contemplating
a lawful universe usually amounts to assuming
that the laws of nature are objective,
have always existed and will exist,
and they are written in the language of mathematics.
Taken~this for granted
is an assumption which raises many problems, some of which will be discussed later.
In~this tradition science can
be done in one way,
the Galileo-Newton one; but if there are no laws,
we~can be freed to pursue  other methodological options, some of which are not entirely unproblematic.
Continuing to enrich
the fundamental Greek practice of scientific observation, thinking and debating on different theoretical interpretations of
phenomena with other  methods, like  the experimental methods (since Galileo) and the mathematical models
(since Descartes and Newton) is obviously desirable.
A step in this direction is to incorporate robust data analytics as a scientific method, see~\cite{miningMD2011,DS2012,doingDS1014,Reed:2015:ECB:2797100.2699414}. However,
suggestions to narrow down the scientific methods to~just a collection of ``empirical evidences'',
to~advance  purely speculative theories (see~\cite{nature_scmeth2014} for physics)
or  to promote the  ``philosophy'' according to which
correlation supersedes causation and theorising (see~\cite{suprious2016}) are~dangerous.\footnote{See the Appendix~\ref{appA} for a more formal discussion.}

The laws governing ``physis'' (nature) and those under which human societies are ruled have often been conflated
and postulated to be of the same origin.
At the dawn of western civilisation Heraclitus held that {\it l\'ogos}\footnote{L\'ogos is the apparent antithesis of {\it x\'aos} in
Hesiod's {\it Theogony}~\cite{hesiod+700-2}.} permeates everything, %plese check for correct ref,
an arrangement common to all things yet incomprehensible to man (\cite{Diels-fdv,Curd-PresocReader}   and~\cite{ki-57}).
However,
there are  crucial differences between these laws. As
Aristotle argued, a law is ``by nature'' if it is justified by appeal to something other than an agreement or a decision; in
contrast, the laws human societies are ruled by are agreed upon in the Agora. While the former laws have been considered  ``absolute'', the
latter are clearly conventional. For example, the laws of movement are natural  in contrast with
the institutional structure of Greek democracy which is the result of human consensus.
In Rhetoric, I.13, Aristotle discusses also the compatibility between
the natural and the conventional laws.
%this intertwining is
a  characteristic of human justice, in~contrast to divine justice. Both these
laws are different from the concept of ``natural law'' developed in the Greek (Aristotle) as well as the Roman
(Cicero) philosophies. In this philosophical sense a ``natural law''
asserts certain rights inherent by virtue of {\it  human nature}.
Endowed by nature---by God or a transcendent source---such a law can be understood universally through human
reason~\cite{natural_law2009}.
Two~typical laws of Aristotelian   ``physis'' are:
(i) Nothing moves unless one pushes it (there must be a `mover' in order to move it).
(ii) Because motion does exist, the above law implies that there must be a self-moved mover, i.e., a `Prime
Mover'.
Finally, according to the definition of ``natural'' found in the Nicomachean Ethics, V.7, God is both a lawgiver
for humans and the governor of nature,
a view which was inherited by Christianity.

\section{Laws and Limit Constructions}

\label{llc}

The scientific revolution grounded the proposal of new laws of nature on observation and iterable experimentation;
sometimes these types of laws were simply
guessed or invented, but nonetheless on the grounds of a ``meaningful'' (physical, theoretical and practical) framework.
For example,
after several experiments, some of which were just imagined,
Galileo and others~\cite{Drake-on-Galilei} proposed the ``law'' of inertia.
This law is a fundamental conservation principle, the conservation of momentum, and is a limit principle since no
physical body actually moves at constant speed along an Euclidean line--a straight line with no thickness.
Yet, by extrapolating from his observations made on the object of bodies as their friction was changed, Galileo was able to
deduce the concept of inertia, and closely analyse what circumstances affect this asymptotic movement: friction and
gravitation.
Thus, by this scientific process of
induction,
deduction, extrapolation
and abduction~\cite{Peirce-cp,sep-abduction,AbductiveInference},
an Aristotelian, God given, absolute, notion of a law of ``physis'' was radically modified.
The advantage of this notion of physical law based on limit principles and symmetry is visible once Newton made the
connection between falling apples and planets: there is no need to be anyone pulling nor pushing the planets   to move them around.
Indeed, Newton's law of gravitation gives the trajectories of any two bodies in inertial movement within a gravitational
field, including apples and planets.
On the one hand it became possible to derive Kepler's trajectories and laws for one sun and one planet from
Newton's law, without the need for a Prime Mover that is constantly pushing.
On the other hand, Newton realised that, with two or more planets, reciprocal interactions destabilise the planets'
trajectories (which later would be recognised as a result of chaotic non-linearity).
He thus assumed the aid of occasional interventions of God in order to assure the stability of the planetary system
\emph{in secula seculorum}: God, through a few sapient touches, was  the only guarantor of the long term
stability of the Solar System~\cite{laskar1994}.
Poincar\'e later confirmed mathematically this deep intuition of Newton on the asymptotic chaos within the Solar
System (see below for more discussion of this).
We should note, however, that this analysis only makes sense in the mathematical continua.
Inertia is conceived as a limit property; moreover, its understanding as a conservation law (of momentum) alongside the
conservation of energy, as a symmetry in the equations (as a result of Noether's theorems relating symmetries to
conserved quantities~\cite{noether1918invariante}), is~based on continuous symmetries: they are invariant with respect
to continuous translations in space or time.
A few years later, Galileo, Boyle and Mariotte proposed another limit law: they traced the isothermal hyperbolas of
pressure and volume for perfect gases.
Of course, actual gases, as a result of friction, gravitation, inter-particle interactions, etc., do not follow this peculiar
conic section;
yet its abstract, algebraicma formulation and its geometric representation, allowed a uniform and general understanding of
the earliest law of thermodynamics.
Principles referring to inexistent ideal trajectories, at the external limit of phenomena, continued to rule knowledge
constructions in physics.
As another example, let us consider Boltzmann's ergodic principle:
{\em In a perfect gas a particle stays in a region of a given space for an amount of time proportional to the volume of
that region.}
Once~again this is an asymptotic principle, as it uniformly holds only at the infinite limit in time.
On~these grounds, Boltzmann's thermodynamic integral that allows the deduction of the second law of thermodynamics
(regarding the increase in entropy) is also formulated as a limit construction (an integral):
it holds only at the infinite limit of the number of particles in the volume of gas.
Can one prove, or at least corroborate,
these asymptotic principles?
There is no way to put oneself or a measurement instrument at these limit conditions and check for Euclidean straight
lines, hyperbolas or behaviour at the asymptotic limit in time.
One may only try to falsify some consequences~\cite{popper-en};
yet, even~in such cases the derivation itself may be wrong, but  not necessarily the principle.
As has already been pointed out by many philosophers, among them Hume, Berkeley, Kant and Schopenhauer,
all we can produce---and this is a crucial point---is {\em scientific knowledge}: we~understand a lot, but not everything,
through these limit principles that unify all movements, all gases, etc., as~specific instances of inexistent movements and
gases.
And, more importantly, as a result we can construct fantastic tools and machines  that work reasonably well -- but not perfectly well, of course---and have radically changed our lives.
With these machines the westerners dominated the world after the scientific revolution, a non trivial consequence
of their science and its ``absolute'' laws.
We are typing, reading and exchanging data in networks of the latest of these inventions, an excellent, but not perfect,
instance of a limit machine--the~Turing machine.
One of the limit principles of these machines is  Turing's
distinction between hardware and software and the identification between program with data that allows abstract,
mathematical styles of programming all the while (almost) disregarding their material~realisation.

Another important consequence was the discovery of limits of computing,
specifically the incomputability of the halting problem,
and more generally the development of theoretical computer science~\cite{Gruska-foc}.\footnote{These limits can be mitigated from a practical point of view with various methods; for example, the halting problem can be solved probabilistically with arbitrarily high precision~\cite{DBLP:journals/computability/CaludeD18}.}
At the same time the abstract character  obscured the role played by physics in computing:
because of the separation between hardware and software, the role of hardware in computation was largely ignored in theoretical computer science,
arguably delaying with a few decades the understanding and development of physics of computation,
reversible computing and quantum computing,~\cite{1402-4896-35-1-021, fr-kn-mar2,mermin-07}.

\section{Order within Disordered Sequences}

\label{2018-wos-sec2}
In intuitive terms, Ramsey theory states that there exists a certain degree of order in all sets/sequences/strings,
regardless of their composition.
Heuristically speaking, this is so because it is impossible for a collection of data
not to have  ``spurious'' correlations, that is,
relational properties among its constituents which are determined only by the size of the data.
The simplest example of such (spurious) correlation is given by the {\em Dirichlet's pigeonhole principle}
stating that $n$ pigeons sitting in $m<n$ holes result in at least one hole being filled with at least two pigeons.
Or in a party of any six people, some three of them are either mutually acquaintances,
or complete strangers to each other~\cite{Greenwood-Gleason-55,Bostwick-1959}.\footnote{In fact,
there is a second trio who are either mutually acquainted or unacquainted~\cite{Bostwick-1959}.}
This seemingly obvious statements can be used to demonstrate unexpected results; for example, the
pigeonhole principle implies that there are two people in Paris who have the same number of hairs on their heads.
The pigeonhole principle is true for at least two pigeons and one whole; the party result needs at least six people.
A common drawback of both results is their non-effectivity: we know that two people in Paris have the same number
of hairs on their heads, but we don't know who they are.

An important result in Ramsey theory is Van der Waerden theorem (see~\cite{graham}) which states that {\em in every
binary sequence at least one of
the two symbols must occur in arithmetical progressions of every
length.}\footnote{If we interpret 0 and 1 as colours,
then the theorem says that in every binary sequence there exist arbitrarily long monochromatic arithmetical
progressions.} The theorem describes a set of arbitrary large strong correlations -- in the sequence $x_1x_2\dots
x_n\dots$ there exist arbitrary large $k, N$ such that equidistant positions $k, k+t, k+2t, \dots k+Nt$ contain the same
element (0 or 1), that is, $x_k = x_{k+t}= x_{k+2t}, \dots =x_{k+Nt}$.\footnote{Again,
the proof is not constructive.} Crucial here is the fact that the property holds true for {\em every} sequence, ordered or
disordered.\footnote{The finite version of  Van der Waerden theorem shows  that the same phenomenon appears in long enough strings. See more
in~\cite{suprious2016}.}
Are these correlations ``spurious''?
According to Oxford Dictionary, {\it spurious} means ``Not being what it purports to be; false or fake. False, although
seeming to be genuine. Based on false ideas or ways of thinking.'' The (dictionary) definition of the word ``spurious'' is
semantic, that is, it depends on an assumed theory: one correlation can be spurious according to one theory, but
meaningful with respect to another one.

Can we give a definition of ``spurious correlation'' which is independent of {\it any} theory? Following~\cite{suprious2016} a
{\it spurious correlation} is defined in a very restrictive way as follows:
{\it a correlation is {\it spurious} if it appears in a randomly generated string/sequence}.
Indeed, in the above sense a spurious correlation
is ``meaningless'' according to any reasonable interpretation  because, by construction, its values have been generated at ``random'', as
all data in the sequence.  As a consequence, such a correlation cannot provide reliable information on future developments of any type of behaviour.
Of course, there are other reasons making a correlation spurious, even within a ``non-random'' string/sequence.  But, are there  correlations as defined above?
Van der Waerden theorem proves that in every sequence there are spurious correlations in the above sense --
they
can be said to ``emerge''.
Therefore, these spurious correlations can also be re-interpreted as ``emerging laws.''
It is important to keep in mind that these ``laws'' are not properties of a particular sequence,---indeed, they exist in  {\em all} sequences as Van der Waerden theorem proves.
How do the spurious correlations manifest themselves in a number world? From the finite version of   Van der Waerden theorem, the more bits of the sequence describing the
number world we can observe, the longer are the lengths of monochromatic arithmetical progressions.
So, once there are (sufficiently many) data, regardless of their intrinsic structure, ``laws from nowhere''
({\it ex nihilo}) emerge.  In what follows we will work only with the above definition of
spurious correlation.

Are these spurious correlations just simple accidents or  more customary phenomena?  We can  answer this question by analysing the ``sizes'' of the sets of random sequences/strings in which   spurious correlations arise.  As our definition of spurious correlation is  independent of any theory, in answering the above questions we will use a model of randomness for sequences and strings provided by algorithmic information theory~\cite{calude:02,DH} which has the same property.

First, how ``large'' is the set of random sequences? If we work with Martin-L\"of random sequences\footnote{A Turing machine with a prefix-free domain is called self-delimiting. A (self-delimiting) Turing machine which can simulate any other  (self-delimiting) Turing machine
is called universal. A sequence  is Martin-L\"of random if there exists a fixed constant such that every finite prefix (string) of the sequence cannot be compressed by a self-delimiting universal Turing machine by more than a constant~\cite{calude:02}.}, then the answer is ``almost all sequences": the probability of a sequence to be Martin-L\"of random is one.\footnote{This holds true even constructively.} This means that {\it the probability that an arbitrary sequence does not have spurious correlations is zero.}\footnote{Probability zero is not the same as impossibility: there exist infinitely many sequences---like the computable ones---which contain no spurious correlations.}

Second, as human access to sequences is limited to their finite  prefixes,   it is necessary to answer the same question for strings: what is the ``size''
of  ``random'' strings? Using
the incompressibility criterion again~\cite{suprious2016},
a string $x$  of length $n$ is  $\alpha$-random
if no Turing machine  can produce  $x$   from an input with less than  $n- \alpha \cdot n$ bits.\footnote{The minimum length  of an input a Turing machine needs to compute a string of length $n$ lies in the interval $(0, n+c)$, where~$c$ is a fixed constant. From this  it follows that $\alpha \in (0,1)$.}
The number of $\alpha$-random   strings $x$ of length $n$  is larger than
$2^{n}\left(1 - 2^{-\alpha \cdot n }\right) +1$,
and
hence, with finitely many exceptions,  it outnumbers  the number of binary strings of length $n$ which are not $\alpha$-random.\footnote{More precisely, when $n\geq 2/\alpha$.}
More interestingly, the probability that a string $x$ of length $n$ is $\alpha$-random
is larger than
$1- 2^{- \alpha \cdot n }+ 2^{-n}$,
an expression which tends exponentially to 1 as $n$ tends to infinity. This means that {\it the probability that an arbitrary string does not have spurious correlations is as close to zero as we wish provided that its length is large enough, that is, excluding finitely many strings.}

Furthermore, the increase of some types of spurious correlations, i.e.,~emergent ``laws'', can be quantified:
Goodman's inequality~\cite{Goodman-1959,Schwenk-1972} yields lower bounds on how
many spurious correlations are observed as a function of the size of data.
Conversely, Pawliuk recently suggested~\cite{Pawliuk-2017} that Goodman's inequality can be
utilised for testing the (null) hypothesis that a dataset is random:
if the bounds are over-satisfied,
the correlations might be not
spurious, and thus the dataset might not be stochastic.
Can we distinguish between meaningful laws and emerging ``laws''? The answer seems to be negative at least from a
computational point of view.

\section{The Emergence of Turing Complete (Universal) Computation}
\label{2018-was-sectcfcs}

In view of the ``quantification'' of
information content~\cite{chaitin3,calude:02}, how could complexity and structures such as universal
computation, evolve even in principle?
The answer to this question is in the algorithmic information content (complexity) of the
number world.

The proof of  Turing
completeness\footnote{A model of computation is Turing complete---sometimes called universal---if it can simulate  a universal Turing machine.}
of the Game of Life provided by Conway
in~(\cite[]{berl:82}, Chapter 25, What Is Life?) is a useful method for exploring
how complex behaviour like Turing completeness can emerge from very simple rules,
in this case, the rules of cellular automata (see more in~\cite{univgamelife}).
With a universal  Turing machine
and all $\alpha$-random strings one can generate {\it all} strings~\cite{calude:02}.

Is this phenomenon also possible for sequences, that is, for  number worlds?
The answer is affirmative.
According  to a theorem by Ku\v{c}era-G\'{a}cs-Hertlinger (\cite[]{calude:02}, p.~179), there effectively exists
a process $F$---which is continuous  computable operator---which generates all sequences from the set of Martin-L\"of random sequences: in other words,  every sequence is the image from $F$  of a Martin-L\"of random sequence.

\section{Is the World Number Computable?}
\label{computable}
Of course,
there exist  infinitely (countable) computable world numbers.

Can we decide whether  the sequence describing a given world number is computable?
Answering this question is probably impossible  both theoretically and empirically.
However, we can answer a simpler variant of the question: What is the probability that a world number is computable?
If we take as probability the Lebesgue measure~\cite{calude:02}, then the answer is zero.\footnote{Again, one should not think that this means that there are no computable world numbers, see Section~\ref{computable}. The~result follows from the fact that the computable sequences form a countable set.}

The above result shows that the probability that a world
number can be generated by an algorithm is zero.
If we weaken the  above requirement and ask about the probability that
there exists an algorithm which generates  infinitely many bits of a world number,
then the answer remains the same: this probability
is nil.
This result follows from a theorem
in algorithmic information theory saying that the complement
of the above set---the set of  bi-immune sequences\footnote{
A sequence  is bi-immune if  its corresponding set of natural numbers nor its complement contain an infinite computably enumerable subset.}
---has probability one~\cite{calude:02}. A~consequence of this fact,
corroborated   by   an extension of the Kochen-Specker theorem
proving value indefiniteness of quantum observables
relative to rather weak physical assumptions~\cite{vi-aeverywhere-2014}, is that
with probability one a  number world is produced by repeatedly measuring of such a value indefinite observable.

\section{Non-Uniform Evolution}
\label{2018-was-sectnue}
Two examples of world numbers are particularly interesting:  Champernowne world number and Chaitin world number. A Champernowne world number in base two is given by the sequence $$C_2=01000110110000010100111001011101110000\dots $$ which consists in the concatenation of all binary strings enumerated in quasi-lexicographical order~\cite{DGC1933}.\footnote{In base 10, $C_{10} = 12345678910111213141516\dots$.}  A a Chaitin world number is given by a Chaitin $\Omega_U$ number (or halting probability),  that
is the probability that the universal self-delimiting Turing machine  $U$ halts~\cite{rtx100200236p}.
Chaitin world numbers ``hold proofs'' for almost all mathematical known results;
such as as Fermat Last Theorem  (in the 400 initial bits),
Goldbach's conjecture,
or important conjectures like Riemann Hypothesis  (in the  2745 bits initial bits) and P vs.~NP (in the 6,495 initial bits;
cf.~\cite{CE2013}.
Both
world numbers are Borel normal in the sense that every binary string $x$ appears in these sequences infinitely many times with the same frequency, namely
$2^{-|x|}$, where $|x|$ is the length of $x$.
In such a world every text---codified in binary---which was written and will be ever written appears infinitely many times and with the same frequency,  which depends only on the length of the text. In particular, any correlation appears in such a world infinitely many times. However, these worlds are also very different: A Champernowne world number is computable, but a Chaitin world number is highly incomputable because it is Martin-L\"of random. As a consequence, while both number worlds  have all possible correlations repeated infinitely many times, the status of those correlations are different: in a Chaitin world number these correlations are spurious (because of its randomness), but in a Champernowne world number they are not (because  its  computability, hence highly non-randomness).

How  an embedded observer would  ``feel''  to live in such a world?  This is a deep question which needs  more study. Here we will make only a few simple remarks (see also~\cite{calude1999islawful}).

First, no observer or rational agent could decide in a finite time whether they live in a Champernowne or Chaitin world.
Second, any observer or rational agent surviving, or at least recording experimental outcomes, a sufficiently long time
will see many of the previously discovered accepted ``laws'' being refuted.

Third, suppose  intrinsic observers embedded into a mathematical universe experience and ``surf''  these number worlds by their interactions with them;
that is, they perceive long successions of initial bits of their defining infinite sequences.
Assume now that  these sequences are  Champernowne or Chaitin  sequences.
Because of the Borel normality of these sequences, the strings surfed by observers are
Borel normal  as finite objects,
that is, they are distributed uniformly up to finite corrections~\cite{DBLP:conf/dlt/Calude93}.
How would intrinsic observers experience such variations?
In one scenario one may speculate that intrinsically such ``interim'' periods of monotony may not count at all; that is,
these will not be operationally recognised as such:
for an embedded observer~\cite{toffoli:79,svozil-94}, the  world number will remain ``dormant'' while the number world remains monotonous.

Another option, maybe even more speculative, is to assume that, as long as the world number allows for a sufficiently wide variety of substrings,
the intrinsic
phenomenology will, through  emerging character of (self-) perception, ``pick'' its own segment or pieces (of numbers) from all the available ones.
Indeed, it might not matter at all for intrinsic perception whether, for instance, the cycle time is altered (reduced, increased),
or whether the lapse of cycles is arbitrarily exchanged or even inverted:
as long as there are still ``sufficient'' patterns and number states emerging could ``process'' and ``use,''
lawfulness and consciousness will always ensue~\cite{permutationcity}.

\section{Is the Universe Lawless?}
\label{lawless}

In this section we add another argument---to many others~\cite{armstrong_1983,vanFraassen1989-VANLAS,calude1999islawful,lawlses_rosen2010,calude2013theeinai,chaos_multiverse2017,Mueller-2017,Cabello-2018-BornRule}---in favour of the hypothesis in the title.

There are uncountable infinite binary sequences, each of which could be a (the) ``true'' simplest model of our universe.
Among these candidates, we have the set of Martin-L\"of random sequences,
which will fit with the hypothesis:
this set is very large, because as we have already mentioned,
%due to a well-known theorem of Martin-L\"of~\cite{calude:02},
the probability that an infinite binary sequence is not Martin-L\"of random is (constructively) zero.
However,~the complement of this set---which has then probability zero---is not only infinite,
but also uncountable and therefore cannot be lightly discarded.

The so-called  physical/universal/natural ``laws'' deal with the infinity, on one hand;
but can be verified only on finitely many cases. What about the situation when a
``true model'' is not a Martin-L\"of random sequence, possibly a highly improbable computable one?

In order to be able to attempt to confirm the ``laws''  in this model we have to surf
the initial bits of the infinite sequence.
How many bits can be surfed?
A possible bound  from below is the number of atoms in the universe which is believed to be less than
$\text{Number}_\text{atoms}=10^{82}$. What is then the probability that
an infinite sequence, thought as a model of our universe, starts
with an $\alpha$-random  string of length $\text{Number}_\text{atoms}$?
In this set there are infinitely many Martin-L\"of random sequences
and a sequence is Martin-L\"of random with probability one, see Section~\ref{computable},
but also infinitely many computable sequences.
The analysis in Section~\ref{2018-wos-sec2}
shows that this probability will be larger than

\[1- 2^{- \alpha \cdot 10^{82} }+2^{-10^{82}},\]

\noindent because this is the probability  that an infinite binary sequence starts with an $\alpha$-random string  of length $\text{Number}_\text{atoms}$.
With this probability---which is infinitesimally close to one---{\it every  choice for our  model of our universe } starts with an
$\alpha$-random  string; consequently, all patterns and correlations which can be verified  in this model  are spurious!

Let us hasten to note that spurious does not mean wrong, not  genuine,  useless.
On the contrary, correlations can be, and many times are,   interesting, useful and give us insight about the working of the universe; they are,
however,  local and not universal.

\section{Conclusions}
\label{2018-was-secsum}
According to Heidegger~\cite{Heidegger-1929},
the most profound and foundational
%(``in den Grund'')
metaphysical issue is to think the existent as the existent
({\em ``das Seiende als das Seiende denken''}).
Here the existent is metaphorically interpreted as
an infinite sequence of bits, a Number World.
Rather than answering the primary question~\cite{Heidegger-1935} of why there
is existence rather than nothingness,
this paper has  been mostly concerned with the formal consequences of existence under
the least amount of extra assumptions~\cite{jaynes}.

As it turns out, existence implies that  an
intrinsic and  sophisticated
mixture of meaningful and (spurious) patterns---possibly interpreted as  ``laws''---can arise from x\'aos. The~emergent  ``laws'' abound,  they can be found almost everywhere.
The axioms in mathematics find their correspondents in the ``laws'' of physics  as a sort of ``l\'ogos'' upon which the respective mathematical universe  is ``created by the formal system''.
By analogy, our own universe might be, possibly deceptively and hallucinatory, be perceived as
based upon such sorts of ``laws'' of physics.
The results in Sections~\ref{2018-wos-sec2} and~\ref{lawless}  have  corroborated the Humeanism view,
later promoted by Exner and to some extend by the young Schr\"odinger, that at least
some physical ``laws'' merely arise  from x\'aos;
a picture which is compatible with  the unresolvable/irreducible
lawless hypothesis.   The  analysis presented in this paper suggests that the ``laws'' discovered in science correspond merely to syntactical correlations, are  local and not universal.

As in biological living systems, the dynamics  described above  is not a matter of stable or unstable equilibrium, but of far from equilibrium processes which  are ``structurally stable''.
This ``duality'' is supported in physics by the  hierarchical layers theory~\cite{anderson:73, Pattee2012}. The~simultaneous structural stability and non-conservative behaviour   in biology,  which  is a blend of stability and instability is due to the coexistence of opposite properties such as
order/disorder and  integration/differentiation~\cite{rand_biol2013}.

Such an active and mindful (some might say self-delusional and projective) approach to order in and purpose of
the universe may  be interpreted in accord with the ancient Greek theogony~\cite{hesiod+700-2} by
saying that {\it l\'ogos}, the Gods and the laws  of the universe,
originate from ``the void,'' or, in a less certain interpretation, from {\it x\'aos.}
Very similar concepts were developed in ancient China probably around the same time as Homerus and Hesiod: the {\it I Ching} utilises relational properties of symbols from sophisticated stochastic
procedures providing inspiration, meaning and advice on what has been understood as divine intent and the way the universe operates.
%\medskip

\vspace{6pt}

%%%%%%%%%%%%%%%%%%%%%%%%%%%%%%%%%%%%%%%%%%
\section{Acknowledgments}
We thank the anonymous referees for useful comments, criticism and suggestions which have significantly improved the paper.
K.~Svozil was supported in part by the
John Templeton Foundation's {\em Randomness and Providence: An Abrahamic Inquiry Project}

\appendix
\section{Causation and Correlation: Two Formal Models}\label{appA}
\unskip
%\subsection{}
%The appendix is an optional section that can contain details and data supplemental to the main text. For example, explanations of experimental details that would disrupt the flow of the main text, but nonetheless remain crucial to understanding and reproducing the research shown; figures of replicates for experiments of which representative data is shown in the main text can be added here if brief, or as Supplementary data. Mathematical proofs of results not central to the paper can be added as an appendix.
%
%\section{}
%All appendix sections must be cited in the main text. In the appendixes, Figures, Tables, etc. should be labeled starting with `A', e.g., Figure A1, Figure A2, etc.

To understand better the difference between causation and correlation we present two simple models. In the first model we have two hypotheses,
$x$ and $y$ which can true or false and we denote by $x \succ y$ the proposition
``$x$ is a cause for $y$'' and by $C(x,y)$ the proposition ``$x$ and $y$ are correlated''.
The~logical representations of the new propositions are enumerated in the following Table~\ref{tableA1}: %Please confirm formatting
Indeed, $x \succ y = 1$ if  $x$ is true, then $y$ is true, that is, $x=y=1$. Note that $x \succ y$ is a ``more restrictive''  operator  than the logical implication which
is true also when $0 \rightarrow  y=1$, for every $y\in\{0,1\}$. We have $C(x,y=1)$
if and only  if both $x$ and $y$ are either true or false, that is, $x=y$. If~follows that  $x \succ y$ implies $C(x,y)=1$, but the converse is false.
%\medskip

\begin{table}[h]
\centering
\caption{Causation versus correlation: a logical model.} \label{tableA1}
\begin{tabular}{cccc}
\hline
\textbf{\emph{x}} & \textbf{\emph{y}} & \boldmath{$x \succ y$} & \textbf{\emph{C(x, y)}}  \\
\hline
0 & 0 & 0 & 1 \\
%\midrule
0&1 &0 &0 \\
%\midrule
1 &0 &0 & 0\\
%\midrule
1 &1 &1 & 1\\
\hline
\end{tabular}
%\bigskip

\end{table}
%\end{center}

\medskip

The second model is
inspired by the Fechner-Machian identification
of causation with functional dependence~\cite{Norton-2003-cafs,Heidelberger-2010}:
suppose that data is represented by two sets $X$ and $Y$.
If $f \colon X \rightarrow Y$ is a function from $X$ to $Y$,
then we denote the graph of $f$ by $G_f =\{(x, f(x)) \in  X \times Y \mid f(x)=y\}$.
A relation $R$ between $X$ and $Y$ is a set $R\subseteq X\times Y$.
We say that $x\in X$ is an $f$-cause for $y\in Y$ if $f(x)=y$ and we write $x \succ_f y$.
The elements $x,y$ are correlated by the relation $R$, in writing, $C_R(x,y)$,  if $(x,y)\in R$.
Assume that $G_f \subset R$; if  $x \succ_f y$, then   $C_R(x,y)$ but  the converse implication is not true.

Both models show that correlation is symmetric, but causation is not.
However, the models above  do not reflect a crucial difference
between causation and correlation: the former contributes to the understanding, in an imperfect way, of the phenomenon,
but the second is just a syntactical observation. Causation invites testing, revision, even abandonment;
correlation is static and
without further analysis could be misleading, see~\cite{Vigen:2015aa,spuriousweb}.

%\bigskip
% \bibliographystyle{abbrv}
% \bibliography{svozil,cris_s,cris}

\end{document}